\documentclass[12pt]{article}
\usepackage{axodraw}
\newcommand{\hs}{\hspace*{0.5cm}}

\newcommand{\be}{\begin{equation}}
\newcommand{\ee}{\end{equation}}
\newcommand{\bea}{\begin{eqnarray}}
\newcommand{\eea}{\end{eqnarray}}
\newcommand{\baa}{\begin{eqnarray*}}
\newcommand{\eaa}{\end{eqnarray*}}
\newcommand{\bary}{\begin{array}}
\newcommand{\eary}{\end{array}}
\newcommand{\bit}{\begin{itemize}}
\newcommand{\eit}{\end{itemize}}

\newcommand{\al}{\alpha}

\begin{document}
\renewcommand{\thefootnote}{\fnsymbol{footnote}}
\begin{center}
{\bf QUARTIC GAUGE BOSON COUPLINGS AND \\TREE UNITARITY IN THE
$\mbox{SU(3)}_C \otimes \mbox{SU(3)}_L
\otimes \mbox{U(1)}_N$ MODELS}\\
\vspace{3cm}

{\bf $ \mbox{D. T. Binh}^{a}, \mbox{D. T. Huong}^a, \mbox{Tr. T.
Huong}^a, \mbox{H. N. Long}^{a}$, and $\mbox{D. V.
Soa}^{b,}\footnote{On leave of absence from Department of Physics,
Hanoi University of Education, Hanoi, Vietnam}$}\\
{\it $^a$ Institute of Physics, NCST, P. O. Box 429, Bo Ho, Hanoi
10000, Vietnam }\\
{\it $^b$ Department of Physics,
Chuo University, Kasuga 1-13-27, Bunkyo, Tokyo 112, Japan}\\

\vspace{0.5cm}

\end{center}
\begin{abstract}
The  quartic gauge boson  couplings in the $\mbox{SU(3)}_C \otimes
\mbox{SU(3)}_L \otimes \mbox{U(1)}_N$ models are presented. We
find that the couplings of four {\it differrent} gauge bosons may
have
{\it unusual Lorentz} structure and the couplings sastify the tree
unitarity requirement at high energy limit. \vspace{0.5cm}
\end{abstract}
\section{Introduction}
\hspace*{0.5cm} Although the standard model (SM) of electroweak
interactions  has been verified to great precision in the recent
years at LEP, SLC and other places, there remain a few unanswered
questions concerning the generation structure of quarks and
leptons. In particular the question of the number of generations
remains open and few progress has been made towards the
understanding of the interrelation between generations. Amongst
the possible extensions beyond the SM, the models based on the
$\mbox{SU(3)}_C \otimes \mbox{SU(3)}_L \otimes \mbox{U(1)}_N$ (3
-- 3 -- 1) gauge group~\cite{svs,pp,fr,flt,hnl} are interesting
from this point of view. They have the following intriguing
features: Firstly, the models are anomaly free only if the number
of generations $N$ is a multiple of three. If further one adds the
condition of QCD asymptotic freedom, which is valid only if the
number of generations of quarks is to be less than five, it
follows that $N$ is equal to 3. The second characteristic of these
models is that one generation of quarks is treated differently
from two others. This could lead to a natural explanation for the
unbalancing heavy top quarks, deviations of $A_b$ from the SM
prediction,...

In the SM, electroweak gauge bosons are introduced to preserve the
local  $ \mbox{SU(2)}_L \otimes \mbox{U(1)}_Y$ symmetry. As a
result, there is a universality among the couplings of the
fermions to the gauge bosons, the three gauge bosons, and the four
gauge bosons. This universality forms the basis of the success of
the SM. So far the fermion-gauge-boson couplings were tested
precisely at various colliders, however the direct measurement of
the self-couplings of the gauge bosons is not precise enough to
test the SM at loop level. The measurements performed at LEP1 have
provided us with an extremely accurate knowledge of the parameters
of the $Z$ gauge boson: its mass, partial widths, and total width.
There even is first evidence that the contributions of gauge-boson
loops to the gauge-boson self-energies are indeed
required~\cite{pzs}. Thus, an indirect confirmation of the
existence of the trilinear gauge couplings (TGC's) has been
obtained. Deviation of non-abelian couplings from expectation
would signal new physics. In addition, tests of the trilinear
couplings aim at checking the non-abelian gauge structure, while
quartic ones will provide important information on the nature of
spontaneous symmetry breaking. In~\cite{ls}
 the TGCs in the minimal model and in the 3 - 3 - 1 model with
 right-handed neutrinos have been presented.
 The TGCs and quartic gauge
couplings (QGCs) in the minimal 3 - 3 - 1 model in a $U_e(1)$
covariant gauge were used in consideration of the static
electromagnetic properties of the $W$ boson~\cite{tar}.

In this paper we  present a {\it complete set} of the QGCs in two
main versions\footnote{For recent proposed 3 -3 -1 models
see~\cite{new}}
 of the 3 - 3 - 1 models. We will show
that the tree unitarity requirement will be satisfied in all
scatterings of
longitudinal components of the vector gauge bosons.\\

\section{Quartic gauge boson couplings in the 3 - 3 - 1 models}
\hspace*{0.5cm}We outline two kinds of 3 -- 3 -- 1 models: the
minimal proposed by  Pisano, Pleitez and Frampton~\cite{pp,fr},
and the model with right-handed neutrinos~\cite{flt}.

\hspace*{0.5cm} {\it  A. The minimal 3 -- 3 -- 1 model}

The model treats the leptons as the  $\mbox{SU(3)}_L$
antitriplet~\cite{fr,dng}
\begin{equation}
f^{a}_L = \left( \begin{array}{c}
               e^a_L\\ -\nu^a_L\\  (e^c)^a_L
               \end{array}  \right) \sim (1, 3^*, 0);\
a = 1, 2, 3.
\label{l}
\end{equation}

Two of the three quark generations transform as triplets and
the third generation is treated differently - in antitriplet:
\begin{equation}
Q_{iL} = \left( \begin{array}{c}
                u_{iL}\\d_{iL}\\ D_{iL}\\
                \end{array}  \right) \sim (3, 3, -\frac{1}{3}),
\label{q}
\end{equation}
\[ u_{iR}\sim (3, 1, 2/3), d_{iR}\sim (3, 1, -1/3),
D_{iR}\sim (3, 1, -4/3),\ i=1,2,\]
\begin{equation}
 Q_{3L} = \left( \begin{array}{c}
                 d_{3L}\\ - u_{3L}\\ T_{L}
                \end{array}  \right) \sim (3, 3^*, 2/3),
\end{equation}
\[ u_{3R}\sim (3, 1, 2/3), d_{3R}\sim (3, 1, -1/3), T_{R}
\sim (3, 1, 5/3).\] At the nine gauge bosons  $W^a (a = 1, 2, ...,
8)$ and $B$ of $\mbox{SU(3)}_L$ and $\mbox{U(1)}_N$, four are
light: photon $A,\  Z$ and $W^\pm$. The remaining five are new
gauge bosons $Z',\ Y^\pm$ and doubly charged bilepton $X^{\pm
\pm}$. They are expressed in terms of $W^a$ and $B$  as
\renewcommand{\thefootnote}{\fnsymbol{footnote}}%
\footnote{The leptons may be assigned a
triplet as in~\cite{pp}, however two models
are mathematically identical.}
\begin{eqnarray}
\sqrt{2}\ W^+_\mu &=& W^1_\mu - iW^2_\mu ,
\sqrt{2}\ Y^+_\mu = W^6_\mu - iW^7_\mu ,\nonumber\\
\sqrt{2}\ X_\mu^{++} &=& W^4_\mu - iW^5_\mu. \label{ww}
\end{eqnarray}
In addition to these, neutral gauge bosons are photon, $Z$
and $Z'$~\cite{dng}:
\begin{eqnarray}
A_\mu  &=& s_W  W_{\mu}^3 + c_W\left(\sqrt{3}\ t_W\ W^8_{\mu}
+\sqrt{1- 3\ t^2_W}\  B_{\mu}\right),\nonumber\\
Z_\mu  &=& c_W  W_{\mu}^3 - s_W\left(\sqrt{3}\ t_W\ W^8_{\mu}
+\sqrt{1- 3\ t^2_W}\  B_{\mu}\right),\nonumber\\
Z'_\mu &=&-\sqrt{1- 3\ t^2_W}\ \ W^8_{\mu} + \sqrt{3}\ t_W\
B_{\mu},  \label{aw}
\end{eqnarray}
where we denoted $s_W \equiv \sin \theta_W,\ c_W \equiv \cos
\theta_W, t_W \equiv \tan \theta_W. $\\
 The {\it physical} states are mixtures of $Z$ and
$Z'$:
\begin{eqnarray}
Z_1  &=&Z\cos\phi - Z'\sin\phi,\nonumber\\
Z_2  &=&Z\sin\phi + Z'\cos\phi,\nonumber
\end{eqnarray}
where $\phi$ is a mixing angle.

The mixing angle has to be very small~\cite{dng} $- 1.6 \times
10^{-2} \le \phi \le 7 \times 10^{-4}$, so that, we can safetly
neglect the mixing. It is interesting to note that in this model
$\sin^2 \theta_W(m_{Z_2})$ should be less than 1/4, and it leads
to $m_{Z_2} \le 3.1$ TeV. Moreover, from the muon decay
experiment~\cite{pdg}, $m_Y$  is found to be at least 230 GeV at
90 \%  CL. The spontaneous symmetry breaking yields a splitting on
the bileptons masses~\cite{lng} \be | M_X^2 - M_Y^2 | \leq 3\
m_W^2. \label{maship} \ee

The quartic couplings arise from
\begin{equation}
{\cal L}_{QGC} =  \frac{g^2}{4}\ f^{abc} \ f_{ade}\ W_{b \mu}\
W_{c \nu} \ W^{d \mu}\ W^{e \nu}.
\end{equation}
Expressing $W^a \ (a = 1, 2, ..., 8)$ in terms of physical fields
thank to Eqs (\ref{ww}) and (\ref{aw}), after straightforward but
tedious calculation we get

\begin{eqnarray}
\frac{1}{g^2}{\cal L}^{min}_{QGC} & =& \frac{1}{2} \left(
W^+.W^-W^+ .W^-- W^+.W^+W^-.W^- \right)+ \frac{1}{2} \left(
Y^+.Y^-Y^+ . Y^- -  Y^+.Y^+Y^-.Y^- \right)  \nonumber \\  &+&
\frac{1}{2}\left(
X^{++}.X^{--}X^{++}.X^{--}-X^{++}.X^{++}X^{--}.X^{--} \right )
\nonumber \\
 &+ &\frac{1}{2}  \left( W^+.W^-Y^+.Y^-
  +  W^+ . Y^+ W^-.Y^- - 2 W^+.Y^-W^-.Y^+\right)  \nonumber \\
  &+&\frac{1}{2}\left(W^+.W^-X^{++}.X^{--}+W^+.X^{--}W^-.X^{++}-
  2 W^+. X^{++} W^- . X^{--} \right)  \nonumber  \\
  &+& \frac{1}{2}\left( Y^+.Y^-X^{++}.X^{--}+Y^+.X^{--}Y^-.X^{++}-
  2Y^+.X^{++}Y^-.X^{--} \right)  \nonumber \\
&-&s_W^2 \left[\left( A.W^+A.W^--A.AW^+.W^-
\right)+\left(A.Y^+A.Y^- - A.AY^+.Y^- \right) \right] \nonumber
\\ &-&4s_W^2\left( A.X^{++}A.X^{--} - A.AX^{++}.X^{--}\right)
 -c_W^2\left(Z.W^+Z.W^- - Z.ZW^+.W^- \right) \nonumber \\
&-&\frac{ \left(c_W-3s_Wt_W\right)^2}{4}\left[ \left(Z.Y^+Z.Y^-
-Z.Z Y^+.Y^-\right) + \left(
Z.X^{++}Z.X^{--}-Z.ZX^{++}.X^{--}\right) \right]  \nonumber \\
&-&\frac{3}{4}\left(1-3 t_W^2 \right) \left[
\left(Z'.Y^+Z'.Y^--Z'.Z'Y^+.Y^- \right) + \left(
Z'.X^{++}Z'.X^{--}-Z'.Z'X^{++}.X^{--} \right) \right] \nonumber \\
&-& c_W s_W \left( A.W^+Z.W^-+A.W^-Z.W^+ -2A.ZW^+.W^- \right)
\nonumber \\
 &+&\frac{1}{2}s_W(c_W+3s_Wt_W)\left(A.Y^+Z.Y^-+
A.Y^-Z.Y^+ - 2A.ZY^+.Y^- \right) \nonumber \\
   &-&s_W(c_W-3s_Wt_W)
   \left(A.X^{++}Z.X^{--}+A.X^{--}Z.X^{++}-2A.ZX^{++}.X^{--} \right)
   \nonumber \\
&+&\frac{1}{2}s_W \sqrt{3(1-3t_W^2)} \left(
A.Y^+Z'.Y^-+A.Y^-Z'.Y^+ -2A.Z'Y^+.Y^- \right)  \nonumber \\
 &+& s_W \sqrt{3(1-3t_W^2)} \left(
 A.X^{++}Z'.X^{--}+A.X^{--}Z'.X^{++}-2A.Z'X^{++}.X^{--} \right)
  \nonumber \\
&-&\frac{1}{4}(c_W+3s_Wt_W)\sqrt{3(1-3t_W^2)}\left(Z.Y^+Z'.Y^-+Z.Y^-Z'.Y^+
-2Z.Z'Y^+.Y^- \right)  \nonumber \\
&+&\frac{1}{4}(c_W-3s_Wt_W)\sqrt{3(1-3t_W^2)}
\left(Z.X^{++}Z'.X^{--}+Z.X^{--}Z'.X^{++} -2 Z.Z'X^{++}.X^{--}
\right)  \nonumber \\
 &+&\frac{1}{4}\sqrt{6(1-3t_W^2)}\left(Z'.Y^+W^+.X^{--}+Z'.X^{--}W^+.Y^+
 -2Z'.W^+Y^+.X^{--}  \right) \nonumber \\
  &+&\frac{3s_W}{\sqrt{2}}\left(
A.W^+Y^+.X^{--}-A.Y^+W^+.X^{--}\right) \nonumber\\
&+&\frac{3}{2 \sqrt{2}} \left[ s_W t_W\left(
Z.Y^+W^+.X^{--}+Z.X^{--}W^+.Y^+-2 Z.W^+Y^+.X^{--}  \right) \right.
 \nonumber \\
&+& \left. c_W \left(Z.X^{--}W^+.Y^+-Z.Y^+W^+.X^{--} \right)
\right]  + h.c
 \label{qgcm}
\end{eqnarray}
where  the following notation was used: $X.Y \equiv X_\mu Y^\mu$.\\

 The vertices in this model are listed in Table 1. In our
 assumption, {\it all} charged boson lines are taken to be entering into
  the vertices. We remind that in the SM  the QGCs
  contain two parts: the first is
coupling strenght $\propto g^2$ and the second is {\it common}
denoted by $S_{\mu\nu,\rho\lambda}$ in the Cheng \& Li
textbook~\cite{cl}.

\begin{table}[htb]\caption{Quartic couplings in the minimal 3 -- 3
-- 1 model}
\end{table}
\begin{center}
\begin{tabular}{|c|c|}  \hline \hline
Vertex & coupling constant/$g^{2}$   \\  \hline \hline $ W^+_\mu
W^-_\nu  W^+_\al W^-_\beta  $  & $ S_{\mu \al,\nu \beta}$\\
  \hline $ Y^+_\mu  Y^-_\nu  Y^+_\al  Y^-_\beta $  &
   $S_{\mu \al,\nu \beta}$ \\
\hline $ X^{++}_\mu  X^{--}_\nu  X^{++}_\al  X^{--}_\beta $  &
 $ S_{\mu \al,\nu \beta}$\\
\hline $ W^+_\mu  W^-_\nu
Y^+_\al   Y^-_\beta $ & $ S_{\mu \beta,\nu \al }/2$\\
 \hline $ W^+_\mu  W^-_\nu  X^{++}_\al X^{--}_\beta$  &
  $  S_{\mu \al, \nu \beta}/2$\\
\hline $ Y^{+}_\mu Y^{-}_\nu X^{++}_\al X^{--}_\beta $ &
 $  S_{\mu \al, \nu \beta}/2 $\\
\hline $\gamma_\mu  \gamma_\nu  W^+_\al  W^-_\beta$ &
 $ - s^{2}_W S_{\mu \nu ,\al \beta} $\\
\hline $\gamma_\mu  \gamma_\nu  Y^+_\al  Y^-_\beta$ &
 $ - s^{2}_W  S_{\mu \nu ,\al \beta}$
\\  \hline $\gamma_\mu  \gamma_\nu  X^{++}_\al  X^{--}_\beta$  &
 $ - 4 s^{2}_W S_{\mu \nu ,\al \beta} $ \\
\hline $ Z_\mu  Z_\nu  W^+_\al  W^-_\beta $  &
 $ - c^{2}_W  S_{\mu \nu ,\al \beta} $\\
\hline $ Z_\mu  Z_\nu  Y^{+}_\al Y^{-}_\beta$ &
$ - (c_W - 3 s_W t_W)^2  S_{\mu \nu ,\al \beta}/4 $ \\
\hline $ Z_\mu  Z_\nu  X^{++}_\al  X^{--}_\beta $  &
 $ - (c_W - 3 s_W t_W)^2 S_{\mu \nu ,\al \beta}/4 $ \\
\hline $ Z'_\mu  Z'_\nu  Y^{+}_\al Y^{-}_\beta$ &
 $ - 3(1- 3t^2_W) S_{\mu \nu ,\al \beta}/4 $\\
\hline $ Z'_\mu  Z'_\nu  X^{++}_\al  X^{--}_\beta $  &
 $ - 3(1-3t^2_W)  S_
 {\mu \nu ,\al \beta}/4 $ \\
  \hline $ \gamma_\mu  Z_\nu  W^+_\al  W^-_\beta$  &
   $ - c_W s_W S_{\mu \nu ,\al \beta} $\\
\hline $ \gamma_\mu  Z_\nu  Y^{+}_\al Y^{-}_\beta$ &
 $  s_W(c_W + 3 s_W t_W) S_{\mu \nu ,\al \beta}/2  $\\
\hline $\gamma_\mu  Z_\nu  X^{++}_\al  X^{--}_\beta $  &
  $ -s_W(c_W - 3 s_W t_W) S_{\mu \nu ,\al \beta} $\\
\hline $ \gamma_\mu  Z'_\nu   Y^{+}_\al  Y^{-}_\beta$ &
 $  s_W\sqrt{(3 - 9 t_W^2)}  S_{\mu \nu ,\al \beta}/2 $\\
  \hline $\gamma_\mu  Z'_\nu  X^{++}_\al  X^{--}_\beta $  &
  $ s_W \sqrt{(3 - 9 t_W^2)} S_{\mu \nu ,\al \beta} $\\
  \hline $ Z_\mu   Z'_\nu  Y^{+}_\al Y^{-}_\beta$ &
  $ - (c_W + 3 s_W t_W)\sqrt{(3 - 9 t_W^2)} S_{\mu \nu ,\al \beta}/4$\\
  \hline $ Z_\mu   Z'_\nu  X^{++}_\al X^{--}_\beta $  &
   $(c_W - 3 s_W t_W) \sqrt{(3 - 9 t_W^2)} S_{\mu \nu ,\al \beta}/4$\\
\hline $ Z'_\mu  W^+_\nu  Y^{+}_\al X^{--}_\beta$ &
 $  \sqrt{6(1-3t_W^2)} S_{\mu \nu ,\al \beta}/4$\\
\hline $\gamma_\mu  W^{+}_\nu Y^{+}_\al X^{--}_\beta$ &
 $ 3 s_W  V_{\mu \nu \al \beta}/\sqrt{2}$\\
\hline $ Z_\mu   W^+_\nu  Y^{+}_\al X^{--}_\beta$ &
 $ 3\left(s_W t_WS_{\mu \nu ,\al \beta }
 +c_W U_{\mu \beta \nu \al}\right)/(2 \sqrt{2}) $\\
\hline
\end{tabular}
\end{center}
Here the following notations were used
 \bea
 S_{\mu \nu,\al \beta}&\equiv&g_{\mu \al }g_{\nu \beta}+g_{\mu
\beta}g_{\nu \al} -2g_{\mu \nu}g_{\al \beta}, \nonumber \\
 V_{\mu \nu \al \beta}& \equiv &g_{\mu \nu}g_{\al \beta}-g_{\mu
 \al}g_{\nu \beta},   \nonumber\\
 U_{\mu \beta \nu \al}&\equiv&  g_{\mu \beta}g_{ \nu \al}-g_{\mu
 \al}g_{\nu \beta}.
\eea

From Table 1, we see that the two last vertices $\gamma W Y X$ and
$Z W Y X$ are not proportional to the usual $S_{\mu \al, \nu
\beta}$, that is why we call it {\it unusual Lorentz structure}.
$S_{\mu \al, \nu \beta}$ is
 symmetric in permutation of $\mu$ and $\al$, and
in permutation of $\nu$ and $\beta$, so the usual quartic gauge
boson vertices (exclusive of two mentioned ones) are {\it
symmetric} in permutation of two particles with {\it the same
electric charges}.

 \hspace*{0.5cm} {\it  B. The model with
right-handed neutrinos}\\
In this model, leptons are in a triplet:
\begin{equation}
f^{a}_L = \left( \begin{array}{c}
               \nu^a_L\\ e^a_L\\ (\nu^c_L)^a
\end{array}  \right) \sim (1, 3, -1/3), e^a_R\sim (1,
1, -1).
\label{l2}
\end{equation}
The first two generations of quarks are in antitriplets while the
third one is in a triplet: \be Q_{iL} = \left( \begin{array}{c}
                d_{iL}\\-u_{iL}\\ D_{iL}\\
                \end{array}  \right) \sim (3, 3^*, 0),
\label{q} \ee
\[ u_{iR}\sim (3, 1, 2/3), d_{iR}\sim (3, 1, -1/3),
D_{iR}\sim (3, 1, -1/3),\ i=1,2,\] \be
 Q_{3L} = \left( \begin{array}{c}
                 u_{3L}\\ d_{3L}\\ T_{L}
                \end{array}  \right) \sim (3, 3, 1/3),
\ee
\[ u_{3R}\sim (3, 1, 2/3), d_{3R}\sim (3, 1, -1/3), T_{R}
\sim (3, 1, 2/3).\] The doubly charged bileptons of the minimal
model are replaced here by complex neutral  ones: \bea \sqrt{2}\
W^+_\mu &=& W^1_\mu - iW^2_\mu ,
\sqrt{2}\ Y^-_\mu = W^6_\mu - iW^7_\mu ,\nonumber\\
\sqrt{2}\ X_\mu^o &=& W^4_\mu - iW^5_\mu. \eea For a  shorthand
notation, hereafter we will use $X^o \equiv X$.

\hs  The {\it physical} neutral gauge bosons are again related to
$Z, Z'$ through the mixing angle $\phi$. Together with the photon,
these are defined as follows~\cite{hnl}:
\begin{eqnarray}
A_\mu  &=& s_W  W_{\mu}^3 + c_W\left(- \frac{t_W}{\sqrt{3}}\
W^8_{\mu} +\sqrt{1-\frac{t^2_W}{3}}\  B_{\mu}\right),
\nonumber\\
Z_\mu  &=&  c_W  W^3_{\mu} - s_W\left( -\frac{t_W}{\sqrt{3}}\
W^8_{\mu}+
\sqrt{1-\frac{t_W^2}{3}}\  B_{\mu}\right),  \\
Z'_\mu &=& \sqrt{1-\frac{t_W^2}{3}}\
W^8_{\mu}+\frac{t_W}{\sqrt{3}}\ B_{\mu}.\nonumber \label{apstat1}
\end{eqnarray}

  The symmetry-breaking hierarchy gives us splitting on the
bileptons masses~\cite{li} \be | M_X^2 - M_Y^2 | \leq m_W^2.
\label{mashipr} \ee Therefore in the future studies it is
acceptable to put $M_X \simeq M_Y$.

The constraint on the $Z - Z'$ mixing based on
the $Z$ decay, is given~\cite{hnl}:
$-2.8 \times 10^{-3} \le \phi \le 1.8 \times 10^{-4}$, and
in this model we have not only a limit for $\sin^2 \theta_W$
but also the {\it upper} limit for new gauge bosons.

From neutrino-electron scattering one gets a lower limit for
$M_{Z_2}$ in the range of 400 GeV, and the muon decay
data~\cite{pdg} gives a lower bound for $Y$ bosons: 230 GeV (90 \%
CL). The symmetry-breaking hierarchy gives us a bilepton mass
splitting: $m_Y \simeq m_X $ (at least at the tree
level)~\cite{li}

 \be | M_X^2 - M_Y^2 | \leq \
m_W^2. \label{mashipr} \ee

The similar cumbersome calculation gives the QGCs in this model:

\begin{eqnarray}
\frac{1}{g^2}{\cal L}^{rhn}_{QGC} &=& \frac{1}{2} \left[ \left(
W^+.W^-W^+.W^- - W^+.W^+W^-.W^- \right)+ \left(Y^+.Y^-Y^+.Y^-
-Y^+.Y^+Y^-.Y^- \right) \right.\nonumber \\
 &+& \left(
X.X^*X.X^*-X.XX^*.X^*\right) + \left(W^+.W^-X.X^*+W^+.XW^-.X^*-2
W^+.X^*W^-.X \right)  \nonumber \\ &+&  \left(
Y^+.Y^-X.X^*+Y^+.X Y^-.X^*-2Y^+.X^*Y^-.X \right)  \nonumber\\
 &+& \left. \left( W^+.W^- X.X^* +W^+.X W^-.X^* -2 W^+.X^*
 W^-.X \right)\right]  \nonumber \\
&-&s^2_W \left[ \left(A.W^+A.W^--A.AW^+.W^- \right)+ \left(
A.Y^+A.Y^--A.A Y^+.Y^- \right) \right] \nonumber\\
  &-&c^2_W\left(Z.W^+Z.W^- -Z.ZW^+.W^- \right) -
  \frac{\cos^2{2 \theta_W}}{4c^2_W}\left( Z.Y^+Z.Y^--Z.ZY^+.Y^-
  \right)  \nonumber \\
 &-& \frac{1}{4 c^2_W} \left( Z.X^* Z.X-Z.Z X.X^* \right) +
  \frac{3-t_W^2}{4} \left[\left( Z'.Y^+Z'.Y^--Z'.Z'Y^+.Y^-
  \right)\right.
   \nonumber \\
&+& \left. \left( Z'.XZ'.X^*-Z'.Z'X.X^*
\right)\right]  \nonumber \\
  &-& c_Ws_W \left(A.W^-Z.W^+ + A.W^+Z.W^--2 A.ZW^+.W^- \right)
  \nonumber \\
  &-&\frac{t_W}{2}\cos{2 \theta_W} \left( A.Y^-Z.Y^+ + A.Y^+Z.Y^-
  -2A.Z Y^+.Y^- \right) \nonumber \\
  &+& \frac{s_W}{2}\sqrt{3-t_W^2}\left(A.Y^-Z'.Y^+ + A.Y^+Z'.Y^--2
  A.Z'Y^+.Y^- \right)  \nonumber \\
&-&\frac{\cos{2 \theta_W}}{4 c_W}\sqrt{3 -t_W^2} \left(
Z.Y^-Z'.Y^+ + Z.Y^+Z'.Y^- -2 Z.Z'Y^+.Y^- \right)  \nonumber \\
 &+& \frac{\sqrt{3-t_W^2}}{4 c_W}
 \left( Z.X^*Z'.X + Z.X Z'.X^*-2 Z.Z'X.X^*
\right)  \nonumber \\
   &-& \frac{s_W}{\sqrt{2}} \left( A.Y^-X^*.W^+ + A.W^+X^*.Y^- -2
   A.X^* W^+.Y^- \right)  \nonumber \\
&-& \sqrt{\frac{(3-t_W^2)}{8}} \left( Z'.X^* W^+.Y^- +
Z'.Y^-X^*.W^+ -2 Z'.W^+X^*.Y^- \right) \nonumber \\
&-&\frac{s_Wt_W}{2\sqrt{2}}\left( Z.X^*W^+.Y^- + Z.Y^- W^+.X^* -2
Z.W^+X^*.Y^- \right)  \nonumber \\
 &+& \frac{c_W}{2\sqrt{2}}\left( Z.Y^- W^+.X^* -Z.X^* W^+.Y^-
 \right)  + h.c\\ \nonumber
 \end{eqnarray}

As before, the QGCs in the considered model are presented in Table
2.

\begin{table}[htb]\caption{Quartic couplings in the  3 -- 3
-- 1 model with right-handed neutrinos}
\end{table}
\begin{center}
\begin{tabular}{|c|c|}  \hline \hline
Vertex & coupling constant/$g^2$   \\  \hline \hline $
W_\mu^+ W_\nu^- W_\al^+ W_\beta^- $  & $S_{\mu\al,\nu\beta}$\\
\hline $ Y_\mu^+ Y_\nu^- Y_\al^+ Y_\beta^- $  &
 $S_{\mu\al,\nu\beta}$\\
\hline $ X_\mu
X_\nu^{*} X_\al X_\beta^{*} $  &  $S_{\mu \al,\nu\beta}$\\
\hline $ W_\mu^+ W_\nu^- Y_\al^+ Y_\beta^- $  &  $
S_{\mu\al,\nu\beta}/2$\\
\hline $ W^+_\mu  W^-_\nu  X_\al  X^*_\beta $  & $
 S_{\mu \al,\nu \beta}/2 $\\
\hline $ Y^{+}_\mu Y^{-}_\nu X_\al X^{*}_\beta
$ & $  S_{\mu \beta,\nu \al}/2$\\
  \hline $\gamma_\mu \gamma_\nu W^+_\al W^-_\beta$  & $ - s^{2}_W
  S_{\mu \nu,\al \beta} $\\
\hline $\gamma_\mu  \gamma_\nu  Y^+_\al  Y^-_\beta $  & $-
s^{2}_W  S_{\mu \nu,\al \beta} $ \\
\hline $ Z_\mu  Z_\nu  W^+_\al  W^-_\beta$  & $  -c^{2}_W
  S_{\mu \nu,\al \beta}  $\\
\hline $ Z_\mu  Z_\nu  Y^{+}_\al Y^{-}_\beta $ & $-
\cos^2{ 2\theta} S_{\mu \nu,\al \beta}/(4 c_W^2)  $ \\
\hline $ Z_\mu  Z_\nu  X_\al  X^{*}_\beta  $ & $-
  S_{\mu \nu,\al \beta}/(4 c_W^2)  $ \\
 \hline $ Z'_\mu Z'_\nu  Y^{+}_\al Y^{-}_\beta $ &  $ (3-t^2_W)
    S_{\mu \nu,\al \beta}/4 $\\
\hline $ Z'_\mu  Z'_\nu  X_\al  X^{*}_\beta  $  & $ (3 -
t^2_W) S_{\mu \nu,\al \beta}/4 $ \\
\hline $ \gamma_\mu  Z_\nu  W^+_\al W^-_\beta $  & $
-s_Wc_W S_{\mu \nu,\al \beta} $\\
\hline $ \gamma_\mu  Z_\nu  Y^{+}_\al Y^{-}_\beta$ &
$ - t_W \cos{2 \theta_W}S_{\mu \nu,\al \beta}/2  $\\
\hline $ \gamma_\mu  Z'_\nu  Y^{+}_\al Y^{-}_\beta$ &
 $  s_W \sqrt{3-t^2_W}S_{\mu \nu , \al \beta}/2 $\\
\hline $ Z_\mu  Z'_\nu  Y^{+}_\al Y^{-}_\beta$ &
 $ - \cos{2 \theta_W}\sqrt{3 - t_W^2}S_{\mu \nu,\al \beta}/(4 c_W)  $\\
\hline $ Z_\mu   Z'_\nu  X_\al  X^{*}_\beta $  &
 $ \sqrt{3 - t_W^2} S_{\mu \nu,\al \beta}/(4 c_W)  $\\
\hline $ Z'_\mu  X^*_\nu  W^+_\al  Y^{-}_\beta  $ &
$-\sqrt{(3-t^2_W)}
S_{\mu \al,\nu \beta}/(2\sqrt{2})  $\\
\hline $ \gamma_\mu X^{*}_\nu W^{+}_\al Y^{-}_\beta $ &
 $- s_W S_{\mu \nu,\al \beta}/\sqrt{2}   $\\
 \hline $ Z_\mu W^{+}_\nu  X^{*}_\al Y^{-}_\beta$
& $ -\left( s_W t_W S_{ \mu \nu, \al \beta}
-c_W V_{\mu \beta \nu \al} \right) /(2\sqrt{2})$\\
\hline
\end{tabular}

 \end{center}
 Our next step is to show that the vertices
given here justify the unitarity requirement. For this purpose we
show that the cancellation over quartic divergences in high energy
scattering of gauge bosons.

\section{Quartic divergence cancellation in high energy limit}
\hspace{0.5cm}In this section we show that the above presented
QGCs satisfy tree unitarity requirement. For our purpose we
consider processes in which two gauge bosons are in the initial
and in the final state
\be V(p_1,m_1) +  V(p_2,m_2) \rightarrow
V(k_1,m_3) + V(k_2,m_4)\label{qt},
\ee
where momentum and mass of
the corresponding particle are put in the bracket. Here $V$ stands
for gauge bosons in these models: $\gamma,Z,W,X,Y$.
 As usual we denote the Mandelstam variables
 \be s=
(p_1+ p_2)^2, t= (p_1 - k_1)^2, u= (p_1 - k_2)^2 \ee which satisfy
the  relation \be s+t+u=\sum_{i=1}^{4}m^2_i.\label{stu} \ee In the
center of mass frame in which ${\bf p}_1=-{\bf p}_2\equiv {\bf p},
{\bf k}_1=-{\bf k}_2\equiv {\bf k}$, the differential cross
section is given by
 \be
\left(\frac{d\sigma}{d\Omega}\right)_{cm} = \frac{|M|^2}{64
\pi^2s} \frac{|\bf{k}|}{|\bf{p}|}S .\label{td}\ee Here $M$ is
shorthand notation for the invariant amplitude $\langle f |
|M||i\rangle$ and $S=\prod_a 1/l_a! $ where $l_a$ is the number of
identical particles of type $a$ in the final state. From
(\ref{td}) it is clear that in the high energy limit $s \gg m_i^2$
the unitarity requests $M \sim {\cal O}(1)$. In other words, the
amplitude $M$ cannot contain the terms proportional to
$\frac{s}{m_i^2}$ or $\frac{s^2}{m_i^4}$. It is known that for
four gauge boson scattering the terms $\propto s^2$ arise only
from purely gauge boson contribution (from Feynman diagrams
(1a)--(1d) in figure 1), while the terms $\propto s$ arise from
both gauge boson contribution and the Higgs and would-be
pseudo-Goldstone boson ones (Fig. 1e -- 1g). The massless vector
bosons have only two components, while the massive ones have
three: two transver and one longitudinal components. The
longitudinal component plays a special role: namely in  the high
energy limit, they give the main contributions. For this reason
the longitudinal components are usually used in estimation of high
energy behaviour of scattering amplitudes. In this paper,
therefore  we shall
consider two cases.\\

\subsection{Scattering of massive gauge bosons}
\hs In the high energy limit, the component of vector of
longitudinal polarization is given \be \epsilon^\mu_L(k) = \frac{
k^\mu}{m} + {\cal O}\left(\frac{m}{k_0}\right)\label{lv}\ee

Since the component of $k^\mu$ are growing as $|{\bf k}|$, the
``bad" behaviour of the amplitude concerns with longitudinal
vector bosons. For this reason hereafter we are only working with
 $\epsilon^\mu_L (k)$.\\
\hs We have checked that tree unitarity is satisfied for {\it all}
possible processes such like in (\ref{qt}). For illustration we
consider the following processes \be Z_L(p_1,m_Z) + W_L^+(p_2,m_W)
\rightarrow X_L^{++}(k_1,m_X) + Y_L^-(k_2,m_Y)\label{qt1}\ee in
the minimal model and \be
 Z_L(p_1,m_Z) +
W_L^+(p_2,m_W) \rightarrow X_L^0(k_1,m_X) +
Y_L^+(k_2,m_Y)\label{qt2}
 \ee
 in the model with right-handed neutrinos. The subscript $L$ added
 to the fields indicates the longitudinal component of the vector
 bosons.

 \hs Now we turn to the process (\ref{qt1}). At the tree level the
 Feynman diagrams are depicted in Figure 1.
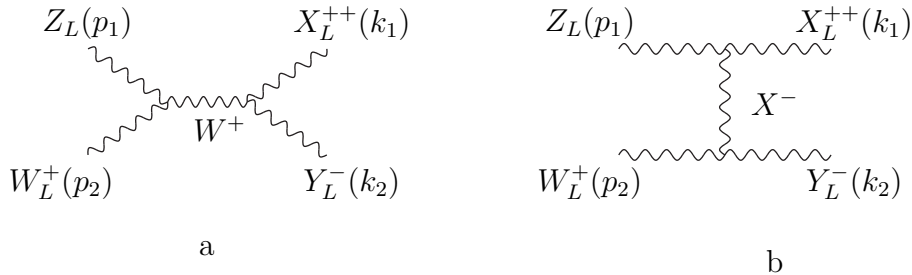
\begin{center}
\begin{picture}(350,100)(0,-20)
\Photon(30,30)(60,50){2}{5}
 \Photon(30,70)(60,50){2}{5}
\Photon(60,50)(90,50){2}{5}
 \Photon(90,50)(120,30){2}{5}
\Photon(90,50)(120,70){2}{5}
 \Text(20,20)[]{$W_L^{+}(p_2)$}
\Text(30,80)[]{$Z_L(p_1)$}
\Text(80,40)[]{$W^+$}
 \Text(130,20)[]{$Y_L^-(k_2)$}
 \Text(130,80)[]{$X_L^{++}(k_1)$}
\Text(75,-5)[]{a}

 \Photon(230,30)(270,30){2}{5}
 \Photon(270,30)(310,30){2}{5}
 \Photon(230,70)(270,70){2}{5}
 \Photon(270,70)(310,70){2}{5}
 \Photon(270,30)(270,70){2}{5}
 \Text(220,20)[]{$W_L^+(p_2)$}
 \Text(220,80)[]{$Z_L(p_1)$}
 \Text(320,20)[]{$Y_L^-(k_2)$}
 \Text(320,80)[]{$X_L^{++}(k_1)$}
 \Text(290,50)[]{$X^-$}
 \Text(290,-10)[]{b}
\end{picture}
\end{center}

\begin{center}
\begin{picture}(350,100)(0,-20)
\Photon(230,30)(310,80){2}{7}
 \Photon(230,80)(310,30){2}{7}

 \Text(220,20)[]{$W_L^+(p_2)$}
 \Text(220,90)[]{$Z_L(p_1)$}
 \Text(320,20)[]{$Y_L^-(k_2)$}
 \Text(320,90)[]{$X_L^{++}(k_1)$}
\Text(270,-5)[]{d}

 \Photon(30,30)(70,30){2}{5}
 \Photon(70,30)(110,70){2}{5}
 \Photon(30,70)(70,70){2}{5}
 \Photon(70,70)(110,30){2}{5}
 \Photon(70,30)(70,70){2}{5}
 \Text(20,20)[]{$W_L^+(p_2)$}
 \Text(20,80)[]{$Z_L(p_1)$}
 \Text(120,20)[]{$Y_L^-(k_2)$}
 \Text(120,80)[]{$X_L^{++}(k_1)$}
 \Text(60,50)[]{$Y^-$}
 \Text(70,-5)[]{c}
\end{picture}
\end{center}

\begin{center}
\begin{picture}(350,120)(0,-20)
\Photon(30,30)(60,50){2}{5}
 \Photon(30,70)(60,50){2}{5}
\DashLine(60,50)(90,50){5}
 \Photon(90,50)(120,30){2}{5}
\Photon(90,50)(120,70){2}{5}
 \Text(20,20)[]{$W_L^+(p_2)$}
\Text(20,80)[]{$Z_L(p_1)$}
 \Text(80,40)[]{$G_W^+$}
 \Text(130,20)[]{$Y_L^-(k_2)$}
 \Text(130,80)[]{$X_L^{++}(k_1)$}
\Text(75,-5)[]{e}

 \Photon(230,30)(270,30){2}{5}
 \Photon(270,30)(310,30){2}{5}
 \Photon(230,70)(270,70){2}{5}
 \Photon(270,70)(310,70){2}{5}
 \DashLine(270,30)(270,70){5}
 \Text(220,20)[]{$W_L^+(p_2)$}
 \Text(220,80)[]{$Z_L(p_1)$}
 \Text(320,20)[]{$Y_L^-(k_2)$}
 \Text(320,80)[]{$X_L^{++}(k_1)$}
 \Text(290,50)[]{$G_X^{++}$}
 \Text(270,-5)[]{f}
\end{picture}
\end{center}

\begin{center}
\begin{picture}(150,150)(0,-80)
\Photon(30,30)(80,30){1.5}{6}
 \Photon(30,80)(80,80){1.5}{6}
\DashLine(80,30)(80,80){5}
 \Photon(80,80)(130,30){1.5}{6}
\Photon(80,30)(130,80){1.5}{6}
 \Text(20,20)[]{$W_L^+(p_2)$}
\Text(20,90)[]{$Z_L(p_1)$}
 \Text(65,60)[]{$G_Y^+$}
 \Text(140,90)[]{$X_L^{++}(k_1)$}
 \Text(140,20)[]{$Y_L^+(k_2)$}
 \Text(80,-5)[]{g}
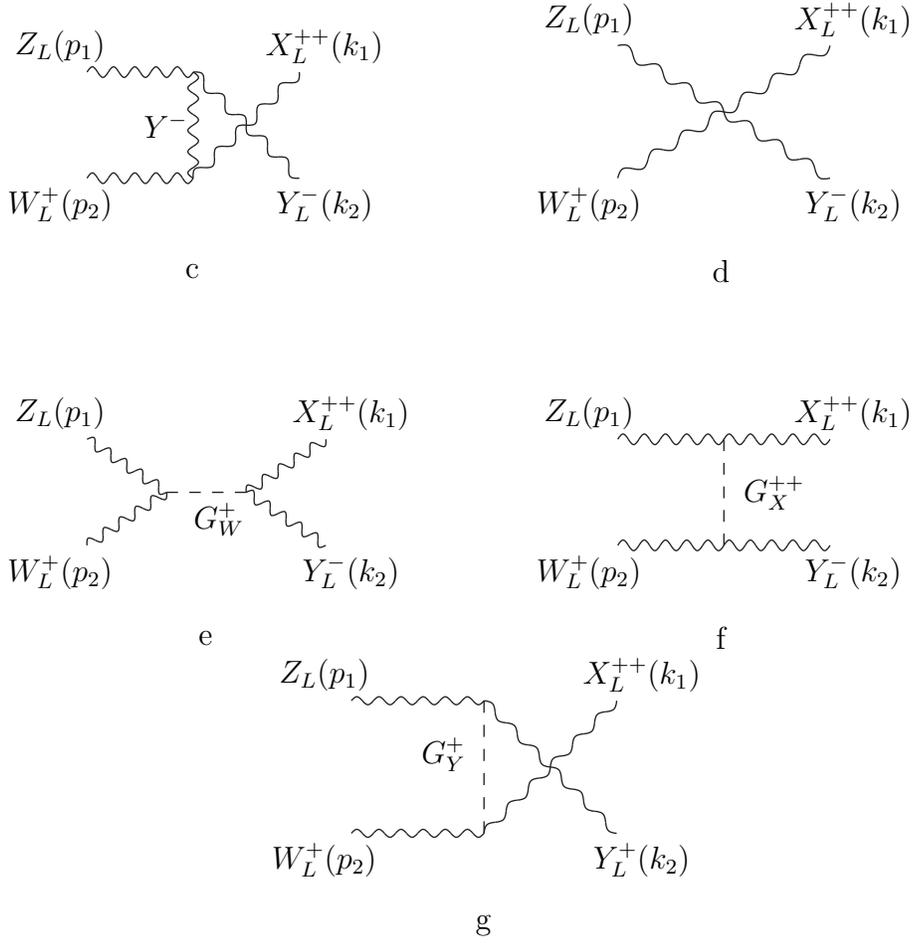
 \Text(70,-40)[]{\mbox{Fig. 1 \hs Tree-level diagrams for the
 process}
  $Z_L W_L^+ \rightarrow X_L^{++} Y_L^-$.}
 \Text(70,-55)[]{Here the wave line represents
  gauge boson and the dashed line - pseudo-Goldstone
  boson  }
  \Text(70,-70)[]{associated with the gauge boson}
\end{picture}
\end{center}

Using the Feynman-t'Hooft gauge, the contributions from Figs. 1a,
1b, 1c and 1d are given, respectively \bea M_{1a}&
=&g_{WWZ}g_{WXY}\frac{(u-t)s}{4m_Zm_Wm_Ym_X}+
{\cal O}\left(\frac{s}{m_i^2}\right)\\
 M_{1b}&=&g_{ZXX}g_{WXY}\frac{(u-s)t}{4m_Zm_Wm_Ym_X} +
 {\cal O}\left(\frac{s}{m_i^2}\right)\\
  M_{1c}&=&g_{ZYY}g_{WXY}\frac{(t-s)u}{4m_Zm_Wm_Ym_X} +
  {\cal O}\left(\frac{s}{m_i^2}\right)\\
   M_{1d}&=&\frac{g_{ZWXY}}{4m_Zm_Wm_Ym_X c_W}\left[
   3s_W^2(2s^2-u^2-t^2) + 3 c_W^2(u^2-t^2)\right] +
  {\cal O}\left(\frac{s}{m_i^2}\right)
  \label{sc} \eea

With the TGCs given in~\cite{ls} (see Appendix, Table 3)   and the
QGCs given in this work we see that the quartic divergences are
indeed vanished.

For process (\ref{qt2}), the Feynman diagrams are depicted in
Fig.2

\begin{center}
\begin{picture}(350,100)(0,-20)
\Photon(30,30)(60,50){2}{5}
 \Photon(30,70)(60,50){2}{5}
\Photon(60,50)(90,50){2}{5}
 \Photon(90,50)(120,30){2}{5}
\Photon(90,50)(120,70){2}{5}
 \Text(20,20)[]{$W_L^{+}(p_2)$}
\Text(30,80)[]{$Z_L(p_1)$} \Text(80,40)[]{$W^+$}
 \Text(130,20)[]{$Y_L^+(k_2)$}
 \Text(130,80)[]{$X_L^0(k_1) $}
\Text(75,-5)[]{a}

 \Photon(230,30)(270,30){2}{5}
 \Photon(270,30)(310,30){2}{5}
 \Photon(230,70)(270,70){2}{5}
 \Photon(270,70)(310,70){2}{5}
 \Photon(270,30)(270,70){2}{5}
 \Text(220,20)[]{$W_L^+(p_2)$}
 \Text(220,80)[]{$Z_L(p_1)$}
 \Text(320,20)[]{$Y_L^+(k_2)$}
 \Text(320,80)[]{$X_L^0(k_1)$}
 \Text(290,50)[]{$X^-$}
 \Text(290,-10)[]{b}
\end{picture}
\end{center}

\begin{center}
\begin{picture}(350,100)(0,-20)
\Photon(230,30)(310,80){2}{7}
 \Photon(230,80)(310,30){2}{7}

 \Text(220,20)[]{$W_L^+(p_2)$}
 \Text(220,90)[]{$Z_L(p_1)$}
 \Text(320,20)[]{$Y_L^+(k_2)$}
 \Text(320,90)[]{$X_L^0(k_1)$}
\Text(270,-5)[]{d}

 \Photon(30,30)(70,30){2}{5}
 \Photon(70,30)(110,70){2}{5}
 \Photon(30,70)(70,70){2}{5}
 \Photon(70,70)(110,30){2}{5}
 \Photon(70,30)(70,70){2}{5}
 \Text(20,20)[]{$W_L^+(p_2)$}
 \Text(20,80)[]{$Z_L(p_1)$}
 \Text(120,20)[]{$Y_L^+(k_2)$}
 \Text(120,80)[]{$X_L^0(k_1)$}
 \Text(60,50)[]{$Y^-$}
 \Text(70,-5)[]{c}
\end{picture}
\end{center}

\begin{center}
+ three diagrams with  inserting pseudo-Goldstone bossons
\end{center}
\begin{center}
\begin{picture}(100,20)(0,0)
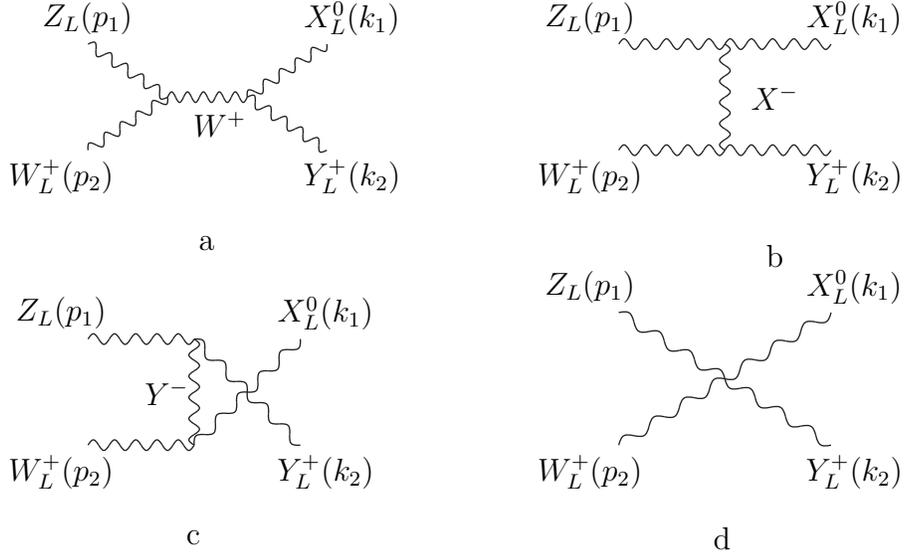
 \Text(30,10)[]{\mbox{Fig. 2 \hs Tree-level diagrams for the
 process}
  $Z_L W_L^+ \rightarrow X_L^{0} Y_L^+$.}
\end{picture}
\end{center}

In exactly the same way, using TGCs in Table 4, it is elementary
exercise to show that  the quartic
divergences are cancelled.\\

\subsection{Processes with massless photon}
 For the process involved photon,
the above manipulation are not applicable. Let us consider in
detail the following process: \be \gamma(p_1,0) +  W_L^+(p_2,m_W)
\rightarrow X_L^{++}(k_1,m_X) + Y_L^-(k_2,m_Y)\label{qt3}\ee The
Feynman diagrams for the above  process are shown in Fig.3

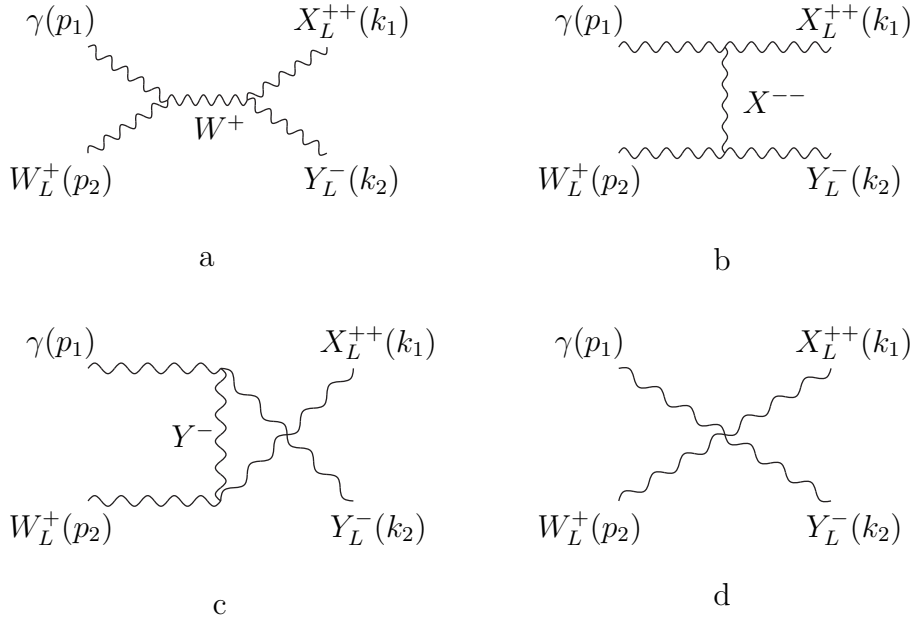
\begin{center}
\begin{picture}(350,120)(0,-20)
\Photon(30,30)(60,50){2}{5}
 \Photon(30,70)(60,50){2}{5}
\Photon(60,50)(90,50){2}{5}
 \Photon(90,50)(120,30){2}{5}
\Photon(90,50)(120,70){2}{5}
 \Text(20,20)[]{$W_L^+(p_2)$}
\Text(20,80)[]{$\gamma(p_1$)}
\Text(80,40)[]{$W^+$}
 \Text(130,20)[]{$Y_L^-(k_2)$}
 \Text(130,80)[]{$X_L^{++}(k_1)$}
\Text(75,-10)[]{a}

 \Photon(230,30)(270,30){2}{5}
 \Photon(270,30)(310,30){2}{5}
 \Photon(230,70)(270,70){2}{5}
 \Photon(270,70)(310,70){2}{5}
 \Photon(270,30)(270,70){1}{5}
 \Text(220,20)[]{$W_L^+(p_2)$}
 \Text(220,80)[]{$\gamma(p_1)$}
 \Text(320,20)[]{$Y_L^{-}(k_2)$}
 \Text(320,80)[]{$X_L^{++}(k_1)$}
 \Text(290,50)[]{$X^{--}$}
 \Text(270,-10)[]{b}
\end{picture}
\end{center}

\begin{center}
\begin{picture}(350,120)(0,-20)
\Photon(230,30)(310,80){2}{7}
 \Photon(230,80)(310,30){2}{7}

 \Text(220,20)[]{$W_L^+(p_2)$}
 \Text(220,90)[]{$\gamma(p_1)$}
 \Text(320,20)[]{$Y_L^-(k_2)$}
 \Text(320,90)[]{$X_L^{++}(k_1)$}
\Text(270,-5)[]{d}

 \Photon(30,30)(80,30){2}{5}
 \Photon(80,30)(130,80){2}{5}
 \Photon(30,80)(80,80){2}{5}
 \Photon(80,80)(130,30){2}{5}
 \Photon(80,30)(80,80){2}{5}
 \Text(20,20)[]{$W_L^+(p_2)$}
 \Text(20,90)[]{$\gamma(p_1)$}
 \Text(140,20)[]{$Y_L^{-}(k_2)$}
 \Text(140,90)[]{$X_L^{++}(k_1)$}
 \Text(70,55)[]{$Y^-$}
 \Text(80,-10)[]{c}
\end{picture}
\end{center}
\begin{center}
+ three  diagrams with  inserting pseudo-Goldstone bossons
\end{center}

\begin{center}
\begin{picture}(100,20)(0,0)
 \Text(30,10)[]{\mbox{Fig. 3 \hs Tree-level diagrams for the
 process}
  $  A  W_L^+ \rightarrow X_L^{++} Y_L^-$.}
\end{picture}
\end{center}

In the Feynman-t'Hooft gauge, the diagrams in Figs. 3a, 3b, 3c
give the following contributions \bea M_{3a}&
=&g_{WW\gamma}g_{WXY}\frac{s[k_1.\epsilon(p_1) -
k_2.\epsilon(p_1)] }{2m_Wm_Ym_X}+
{\cal O}(1)\nonumber\\
 M_{3b}&
=&-g_{XX\gamma}g_{WXY}\frac{t[k_2.\epsilon(p_1) +
p_2.\epsilon(p_1)] }{2m_Wm_Ym_X}+
{\cal O}(1)\nonumber\\
 M_{3c}&
=&-g_{YY\gamma}g_{WXY}\frac{u[k_1.\epsilon(p_1) +
p_2.\epsilon(p_1)] }{2m_Wm_Ym_X}+ {\cal O}(1)\nonumber\\
 M_{3d}&
=&g_{\gamma W X Y}\frac{[u k_2.\epsilon(p_1) + s
p_2.\epsilon(p_1)]}{2m_Wm_Ym_X}+ {\cal O}(1).
  \label{scp} \eea
Noting that $t= - (s + u)$ and using momentum conservation we see
that the leading divergence is cancelled
\[ M_{3a}+  M_{3b}+  M_{3c}+  M_{3d} = 0.\]
With the above manipulation one can check that vertices of gauge
boson self-interaction  satisfy unitarity requirment at the tree
level.\\

\section{Conclusions}
\hspace*{0.5cm}In this paper we have presented a complete set of
quartic gauge boson couplings in both 3 - 3 -1
models. We have shown
 there exist four-{\it different} gauge
boson couplings with unusual Lorentz structure. By consideration
of scattering of longitudinal vector bosons and as well as the
massless photons we have deduced that at the tree level  the
quartic divergences are cancelled and then the unitarity is
satisfied. It is worth to mention that QGCs which contain the
interactions between the SM gauge bosons and the bileptons in the
minimal 3 - 3 -1 model were presented in~\cite{tar}, and their
results are consistent with ours. Other couplings were not
presented there. We hope that the results in this paper are useful
for anyone interested in studying processes involving these
couplings.\\

One of the authors (D. V. S.) expresses sincere gratitude to the
Nishina Memorial foundation for financial support. He would like
also to thank Professor T. Inami and Department of Physics, Chuo
University
for warm hospitality during his visit as a Nishina fellow.
This work was supported in part by National Council for
Natural Sciences of Vietnam.\\

\begin{center}
 {\bf Appendix}
\end{center}
 \hs In this appendix we rewrite trilinear gauge boson couplings (TGCs)
  in the 3 - 3 - 1 models~\cite{ls}.
\vspace*{0.5cm}

\begin{table}[htb]\caption{Trilinear couplings in the minimal
3 -- 3 -- 1 model.}
\end{table}
\begin{center}
\begin{tabular}{|c|c|}  \hline \hline
Vertex & coupling constant/e   \\  \hline \hline $\gamma W^+ W^-$
& 1\\  \hline $Z W^+ W^-$ &$1/t_W$ \\ \hline $\gamma Y^+
Y^-$&$1$\\ \hline $Z Y^+ Y^-$ & $- (1 + 2 s_W^2)/\sin 2\theta_W$\\
\hline $\gamma X^{++} X^{--}$&$ 2 $\\ \hline $Z X^{++} X^{--}$&$(1
- 4 s_W^2)/\sin 2\theta_W$\\ \hline $Z' Y^+ Y^-$&$ - \sqrt{3(1 - 4
t_W^2)}/(2s_W)$\\ \hline $Z' X^{++} X^{--}$&$ - \sqrt{3(1
- 4 t_W^2)}/(2s_W)$\\ \hline $X^{--} Y^+ W^+$&$ 1/(\sqrt{2}\ s_W)$\\
\hline $X^{++} W^- Y^-$&$ 1/(\sqrt{2} \ s_W)$\\ \hline
\end{tabular}\\[2pt]
\end{center}
 \vspace*{0.5cm}

\begin{table}[htb]\caption{Trilinear couplings in the 3 -- 3 -- 1
model with RH neutrinos.}
\end{table}
\begin{center}
\begin{tabular}{|c|c|}  \hline \hline
Vertex & coupling constant/e   \\  \hline \hline $\gamma W^+ W^-$
& $1$\\  \hline $Z W^+ W^-$ &$ 1/t_W$ \\ \hline $\gamma Y^+
Y^-$&$1$\\ \hline $Z Y^+ Y^-$ & $ 1 /  \tan 2\theta_W$\\  \hline
$Z X  X^*$&$- 1 /\sin 2\theta_W$\\ \hline $Z' Y^+ Y^-$&$ -
\sqrt{(3 - t_W^2)}/(2s_W)$\\ \hline $Z' X X^*$&$ - \sqrt{(3 -
t_W^2)}/(2s_W)$\\
\hline $X  W^- Y^+ $&$
1/(\sqrt{2}\  s_W)$\\ \hline $X^* Y^- W^+$&$ 1/(\sqrt{2} \ s_W)$\\
\hline
\end{tabular}\\[2pt]
\end{center}


\vspace*{2cm}


\begin{thebibliography}{99}
\bibitem{svs}M. Singer, J. W. F. Valle and J. Schechter,
Phys. Rev.  D22 (1980) 738;\\  J. W. F. Valle and M. Singer, Phys.
Rev. D28 (1983) 540.
\bibitem{pp} F. Pisano and V. Pleitez, Phys. Rev.
D46 (1992) 410;\\ R. Foot, O.F. Hernandez, F. Pisano and V.
Pleitez, Phys. Rev. D47 (1993) 4158.
\bibitem{fr}P. H. Frampton, Phys. Rev. Lett.
 69, (1992) 2889.
\bibitem{flt} R. Foot, H. N. Long and Tuan A. Tran,
Phys. Rev. D50 (1994) R34.
\bibitem{hnl} H. N. Long, Phys. Rev.
D54 (1996) 4691.
\bibitem{pzs}P. Gambino and A. Sirlin, Phys. Rev. Lett.
 73 (1994) 621; Z. Hioki, Phys. Lett. B340 (1994) 181;
S. Dittmaier {\it et al}, Nucl. Phys. B426 (1994) 249, {\it ibid}
B446 (1995) 334.
\bibitem{ls}H. N. Long and D. V. Soa, Nucl. Phys. B601 (2001) 361.
\bibitem{tar}G. Tavares-Velasco and J. J. Toscano, Phys. Rev. D65
(2001) 013005.
\bibitem{new}R. A. Diaz, R. Martinez, and J. A. Rodriguez,
[hep-ph/0208176]; R. Martinez, W. A. Ponce, and L. A. Sanchez,
Phys. Rev. D64, 075013 (2001);  {\it ibid}  D65, 055013 (2002); Y.
Okamoto and M. Yasue, Phys. Lett. B466, 267 (1999); T. Kitabayshi
and M. Yasue, [hep-ph/0209294].
\bibitem{lng}J. T. Liu and D. Ng, Z. Phys. C62 (1994) 693;
N. A. Ky, H. N. Long, and D. V. Soa, Phys. Lett. B486, 140 (2000).
\bibitem{li} H. N. Long and T. Inami, Phys. Rev. D61 (2000) 075002.
\bibitem{dng}D. Ng, Phys. Rev.  D49 (1994) 4805.
\bibitem{cl}T. P. Cheng and L. F. Li, {\it Gauge theory of
elementary particle physics}, Clarendon press, 1984.
\bibitem{pdg}Particle Data Group, Euro. Phys. J. C15, 105 (2000).
\end{thebibliography}
\end{document}